\documentclass[prb,superscriptaddress,floatfix,twocolumn,amsfonts,amsmath,amssymb]{revtex4}
\usepackage{graphicx} 
\usepackage{dcolumn} 
\usepackage{bm} 
\usepackage{color}
\usepackage{float}

\begin{document}

\title{La$_{2-x}$Ba$_x$CuO$_4$ ($x=\frac{1}{8}$) $\mu$SR data are inconsistent
  with spin stripe but consistent with spin spiral
}

\author{O. P. Sushkov}
\affiliation{School of Physics, University of New South Wales, Sydney 2052, Australia}
\date{\today}

\begin{abstract}
  I analyze available $\mu$SR data and show that it is  inconsistent
  with the spin stripe but consistent with the coplanar spin spiral.
  The plane of the spiral coincides with the CuO$_2$-plane.
  The static expectation value of the spin is $S=0.37\times\frac{1}{2}$.
  \end{abstract}

\maketitle

It is well established that incommensurate magnetic neutron scattering is
a generic phenomenon in cuprate superconductors.
All cuprate families originate from parent Mott insulators that are
collinear antiferromagnets (AF) with spin S=1/2. In neutron scattering the
antiferromagnetism is observed as an elastic peak at the momentum transfer
$Q=(\pi/a,\pi/a)$, where $a$ is the 2D lattice spacing of the  CuO$_2$-plane.
The compounds become conducting and ultimately
superconducting with doping by holes or electrons.
With doping by holes the neutron scattering peak is shifted from
it's AF position by an amount $q$. This is the experimental
manifestation of the spin structure generally incommensurate with the crystal
lattice.
The phenomenon had been observed in single layer cuprates
La$_{1.875}$Ba$_{0.125}$CuO$_4$ (LBCO), Refs.~\cite{Tranquada1995,Tranquada1996}, and
La$_{2-x}$Sr$_{x}$CuO$_4$ (LSCO), Refs.~\cite{Yamada1998,Fujita2002}, in double layer
YBCO, Refs.~\cite{Hinkov2004,Hinkov2008}, and in other cuprates.
The incommensurate scattering can be
elastic/quasielastic~\cite{Tranquada1995,Tranquada1996,Yamada1998,Fujita2002,Hinkov2008} or fully dynamic~\cite{Hinkov2004}. 
In single layer cuprates LBSO and LSCO the incommensurate wave vector $q$
scales linearly with doping x, $q = \frac{2\pi}{a}x$, for $x < 0.16$
(Yamada plot, Ref.~\cite{Fujita2002}). 
The compound La$_{1.875}$Ba$_{0.125}$CuO$_4$ plays a special role. In this case
$x=1/8$, hence generally incommensurate spin structure becomes commensurate and
in LTT phase spins are pinned to the lattice. The pinning reduces fluctuations
and the electron spin structure becomes truly static.
Hence, the electron spin structure can be investigated via muon spin relaxation,
$\mu$SR. In this work I reanalyze  $\mu$SR data~\cite{Keller2014} for
La$_{1.875}$Ba$_{0.125}$CuO$_4$ to shed light on
the spin  structure of underdoped cuprates.

\begin{figure}
    \includegraphics[angle = 0, width=0.42\textwidth]{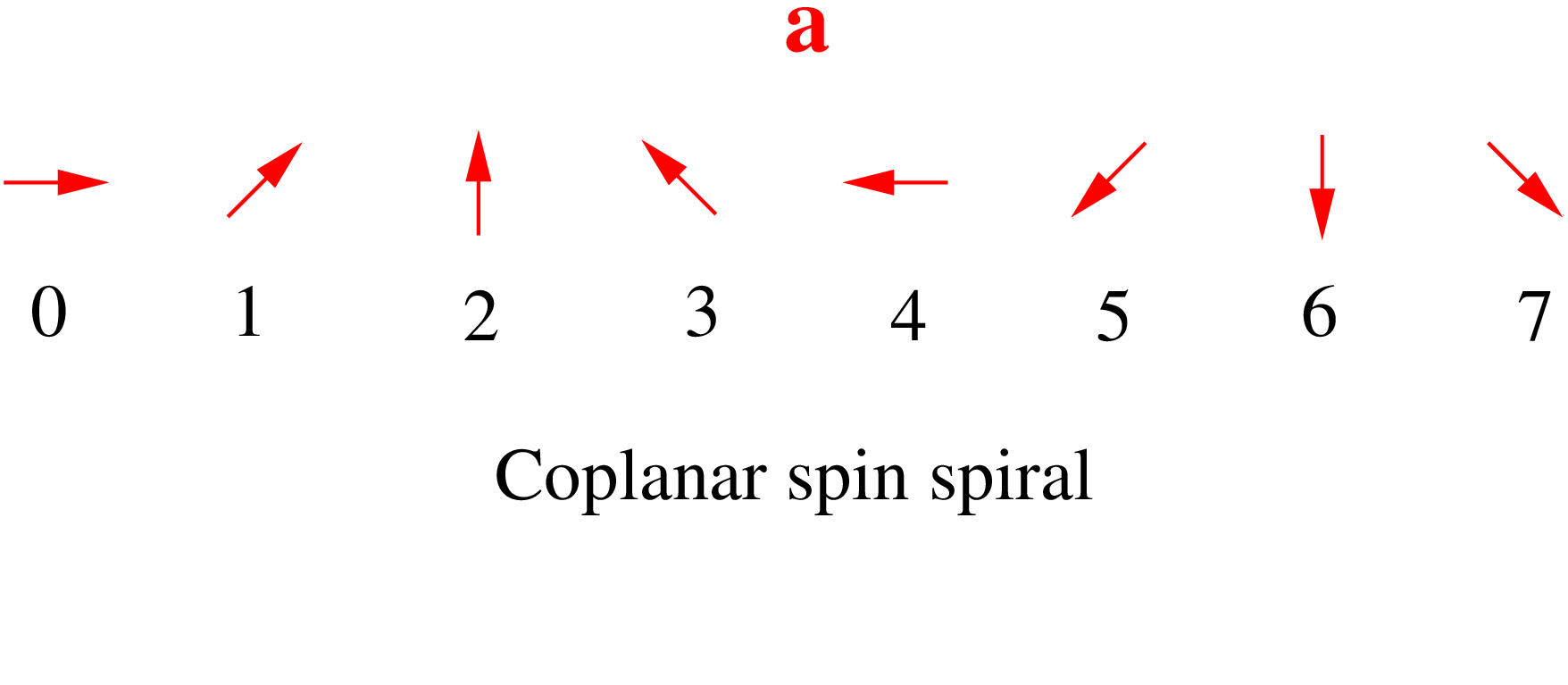}
    \includegraphics[angle = 0, width=0.4\textwidth]{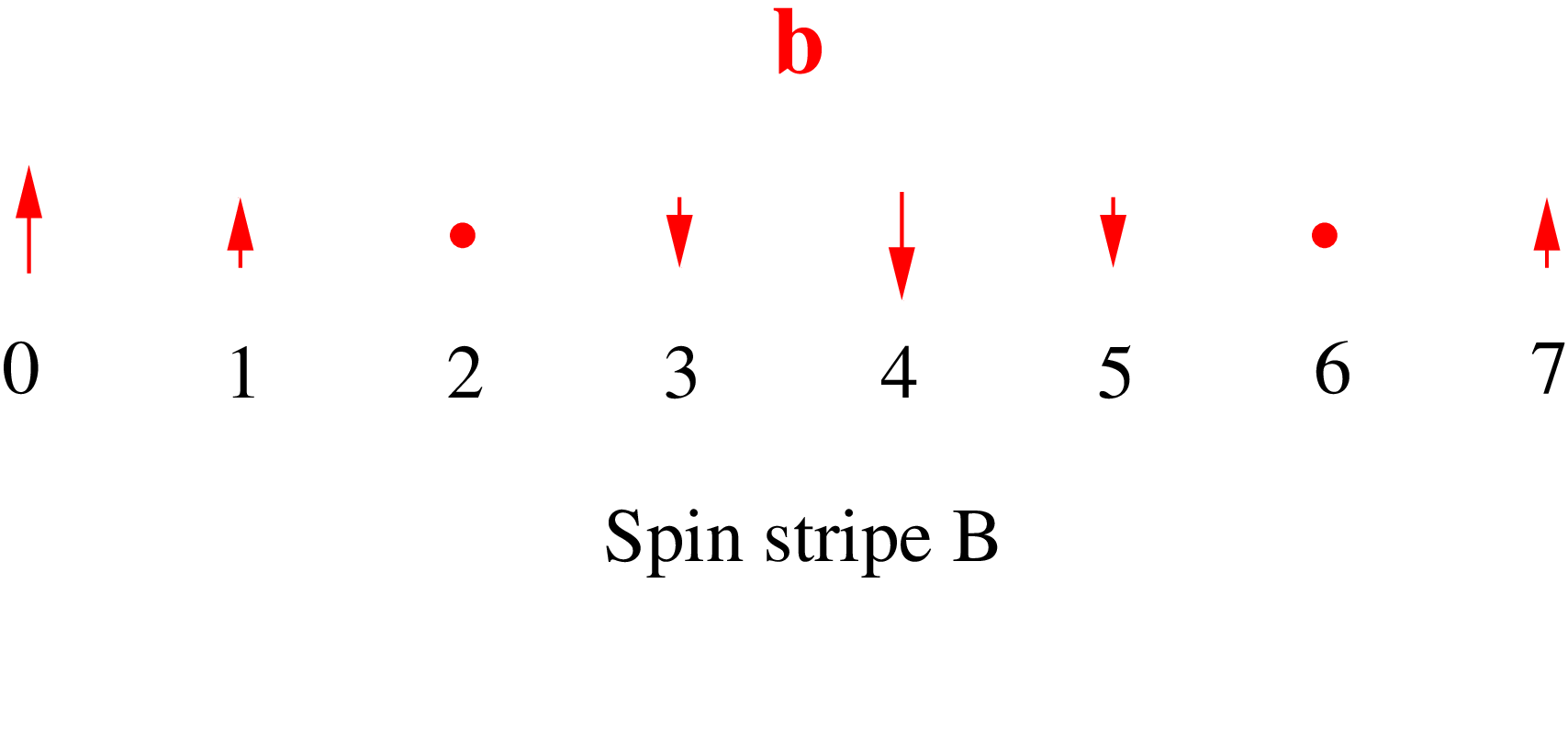}
    \includegraphics[angle = 0, width=0.4\textwidth]{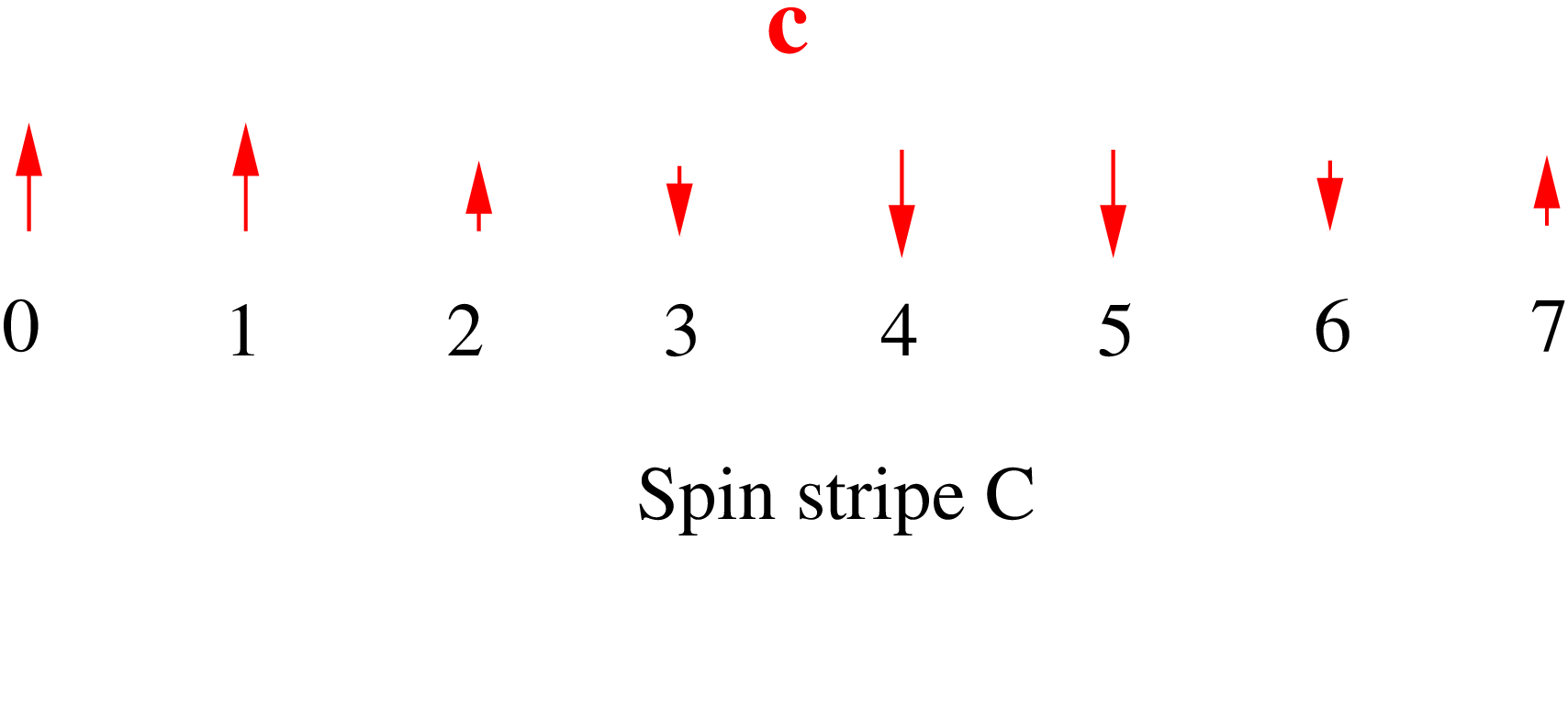}
    \caption{The spin spiral and different types of  spin stripes, the
      numbers enumerate lattice sites. Arrows show directions of the staggered
      spin (every second spin is flipped up side down). 
    }
  \label{FSS}
\end{figure}

There are two alternative ways to explain the incommensurate spin structure,
(i) the coplanar spin spiral and (ii) the spin stripe.
It is convenient to use the  staggered spin to describe the structures:
every second spin is flipped up side down to exclude trivial antiferromagnetic
ordering. Below I use the staggered spin notations.
Equations describing the amplitude and the direction of the static component of the spin
are \begin{eqnarray}
  \label{spineq}
  &&spiral:\ \  {\bf S}\propto {\hat{\bf x}}\cos(q x)+
  {\hat{\bf y}}\cos(q x)  \nonumber\\
  &&stripe:\ \  {\bf S}\propto {\hat{\bf x}}\cos(q x +\varphi)
\end{eqnarray}
In the case of stripes there are always higher Fourier harmonics,
but theoretically they are always small, see e.g. Refs.~\cite{Lorenzana2002,Scholle2023}
The coplanar spin spiral for $q=2\pi/8$ is shown in panel a of Fig.\ref{FSS}.
The spin stripe for $q=2\pi/8$ and $\varphi=0$,
\begin{eqnarray}
  \label{str1}
  S_n\propto \cos\left(\frac{2\pi n}{8}\right), \ \ n=0,1,2,3,4,5,6,7
\end{eqnarray}
is shown in panel b of Fig.\ref{FSS}. I call it ``spin stripe B''.
This is the most commonly considered type of the spin spiral.
For completeness I also consider the spin stripe with $q=2\pi/8$ and
$\varphi=-\pi/8$,
\begin{eqnarray}
  \label{str2}
  S_n\propto \cos\left(\frac{2\pi n}{8}-\pi/8\right), \ \ n=0,1,2,3,4,5,6,7
\end{eqnarray}
The corresponding picture is shown in panel c of Fig.\ref{FSS}. I call it ``spin stripe C''

It is important to note that muons stop at the apical oxygens,
see e.g. Ref.~\cite{Amato2003,Ramadhan2022} Therefore, the depolarization rate is
directly  proportional to the static electron spin located on Cu ion
straight underneath the apical oxygen.

\begin{figure}
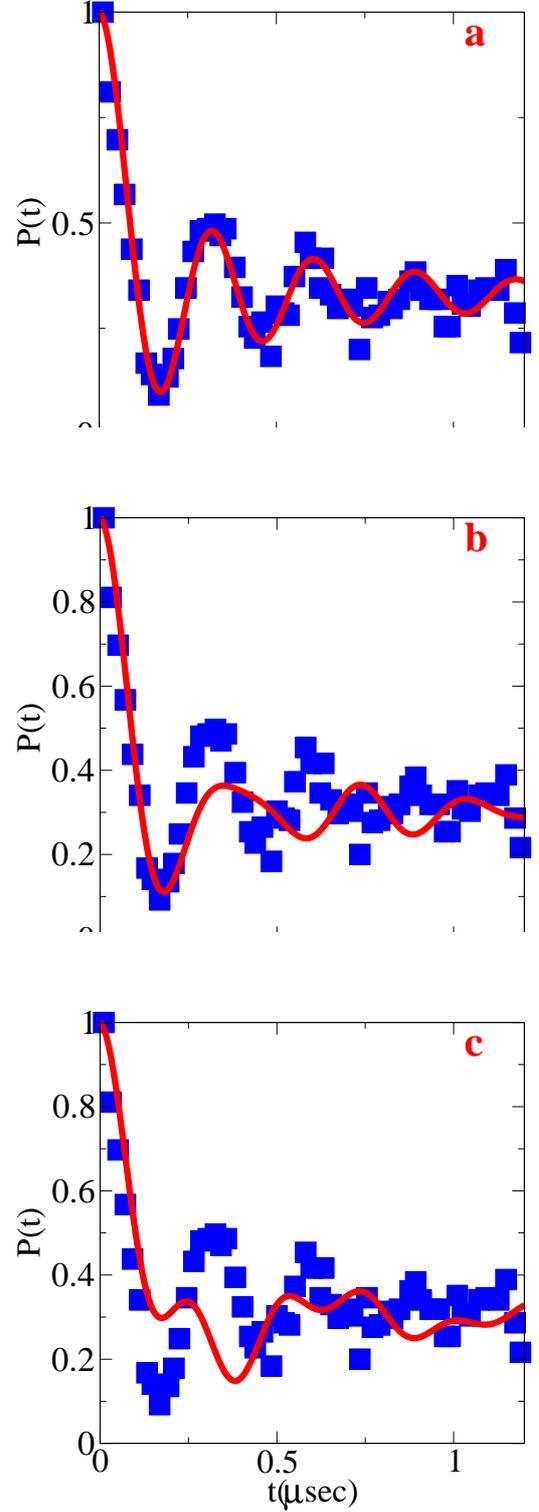

	\begin{center}
		\includegraphics[angle = 0, width=0.4\textwidth]{plot1.eps}
		\includegraphics[angle = 0, width=0.4\textwidth]{plot2.eps}
		\includegraphics[angle = 0, width=0.4\textwidth]{plot3.eps}
		\caption{Muon polarization vs time.
                  Blue squares in all panels show T=5K data form
                  Ref.\cite{Keller2014}
                  The red curve in panel a shows the best fit of data in
                  the spin-spiral model.
                  The red curve in panel b shows the best fit of data in
                  the ``spin-stripe B'' model.
                  The red curve in panel c shows the best fit of data in
                  the ``spin-stripe C'' model.
}
		\label{fit}
	\end{center}
\end{figure}

{\bf Spin spiral.} $\mu$SR data (muon polarization versus time) for La$_{2-x}$Ba$_x$CuO$_4$
($x=\frac{1}{8}$) from Ref.\cite{Keller2014} are presented in all panels of
Fig.\ref{fit} by blue squares.
The fit of data assuming a single depolarization magnetic field using the
standard formula
\begin{eqnarray}
\label{1f}
P(t)=\frac{V_m}{3}\left[2e^{-\lambda_Tt}J_0(\omega t)+e^{-\lambda_Lt}\right]+(1-V_m)e^{-\lambda_{nm}t}
\end{eqnarray}  
has been performed in the original experimental work \cite{Keller2014}.
Here $\omega=2\mu_\mu B$, $\mu_\mu$ is magnetic moment of muon and $B$ is the
field due to the electron spin; $\lambda_T,\lambda_L,\lambda_{nm}$ are
various relaxation rates; and $V_m$ is the relative fraction of muons that
stopped at the sites sensitive to B (stopped at apical oxygens).
Practically, for fitting properties, Eq.(\ref{1f})
can be simplified as
\begin{eqnarray}
\label{1fs}
P(t)={U_m}e^{-\lambda_T t}J_0(\omega t)+(1-{U_m})e^{-\lambda^{\prime}t}
\end{eqnarray}  
Moreover, for the fit within the relevant time range $0 < t < 1.2\mu$sec 
I set $\lambda^\prime=0$, because practically it does not influence the fits.
The fit of the data gives the following parameters,
$\omega=22\times 10^6sec^{-1}$, $\lambda_T=0.9\times 10^6sec^{-1}$,
${U_m}=0.67$. This  results in the red solid curve in panel a of
Fig.(\ref{fit}) that is practically identical to the fit in
Ref.\cite{Keller2014} Agreement with data is excellent.

The fit with Eq.(\ref{1f}) or Eq.(\ref{1fs}) assumes a single depolarization magnetic
field.
The muon sensitive to the magnetic moment of Cu is located on the apical oxygen.
Therefore, the single depolarization field implies that  magnetic moments
at all sites are the same. We also know from neutron scattering that there is
an incommensurate antiferromagnetic structure with period 8 lattice spacing.
Therefore, the single depolarization field calculation is consistent with
the spin spiral shown in Fig.\ref{FSS}. Moreover, the plane of spiral must
coincide with the CuO$_2$-plane, otherwise the depolarization rate at
different sites would be different.

\begin{figure}
  \begin{center}
    \vspace{20pt}
		\includegraphics[angle = 0, width=0.4\textwidth]{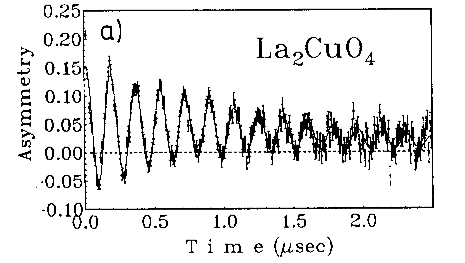}
		\caption{Muon spin relaxation in parent La$_2$CuO$_4$,
                  Ref.\cite{Budnick1987}
}
		\label{pLCO}
	\end{center}
\end{figure}

Fig.\ref{pLCO} shows $\mu$SR relaxation data\cite{Budnick1987} in the parent
La$_2$CuO$_4$. Fitting with  (\ref{1f}) or (\ref{1fs}) gives
$\omega=36\times 10^6sec^{-1}$. This has to be compared with
$\omega=22\times 10^6sec^{-1}$ obtained for La$_{1.875}$Ba$_{0.125}$CuO$_4$.
It is known that the electron magnetic moment in La$_2$CuO$_4$ is
$\mu_e=(0.6\pm 0.05)\mu_B$, Ref.\cite{Yamada1987}
Hence the electron magnetic moment in La$_{1.875}$Ba$_{0.125}$CuO$_4$. is
\begin{eqnarray}
  \label{mue}
\mu_e=\frac{22}{36}\times 0.6\mu_B=0.37\mu_B \ .
\end{eqnarray}

{\bf Spin stripe B model.}
In the spin stripe B, described in Fig.\ref{FSS}b and  Eq.(\ref{str1}),
the depolarization magnetic field at sites with n=1,3,5,7
is by factor $1/\sqrt{2}$ smaller than that at sites with n=0,4.
Therefore, to describe  the muon spin relaxation one has to perform the
following replacement in Eqs.(\ref{1f}),(\ref{1fs}) 
\begin{eqnarray}
\label{repl}
J_0(\omega t) &\to& \frac{1}{3}[J_0(\omega t)+2J_0(p\omega t)]\nonumber\\
p&=&1/\sqrt{2}=0.71
\end{eqnarray}
Here $\omega$ corresponds to relaxation at sites n=0,4.
The best fit achieved at
$\omega=26\times 10^6sec^{-1}$, $\lambda_T=1.0\times 10^6sec^{-1}$,
${U_m}=0.7$ is shown by solid red
line in panel b of Fig.\ref{fit}. The fit c is not consistent with data.
This  rules out the spin-stripe B model.

{\bf Spin stripe C model.}
In the spin stripe C, described by Fig.\ref{FSS}c and Eq.(\ref{str2}),
the depolarization magnetic field at sites with n=2,3,6,7
is by factor $\cos(3\pi/8)/\cos(\pi/8)$  smaller than that at sites with
n=0,1,4,5.
Therefore, to describe  the muon spin relaxation one has to perform the
following replacement in Eqs.(\ref{1f}),(\ref{1fs}) 
\begin{eqnarray}
\label{repl1}
J_0(\omega t) &\to& \frac{1}{2}[J_0(\omega t)+J_0(p\omega t)]\nonumber\\
p&=&\frac{\cos(3\pi/8)}{\cos(\pi/8)}=0.41
\end{eqnarray}
Here $\omega$ corresponds to relaxation at sites n=0,1,4,5.
The best fit achieved at
$\omega=26\times 10^6sec^{-1}$, $\lambda_T=1.0\times 10^6sec^{-1}$,
${U_m}=0.7$ is shown by solid red
line in panel c of Fig.\ref{fit}. The fit is inconsistent with data.
This rules out the ``spin-stripe C'' model.

\begin{figure}
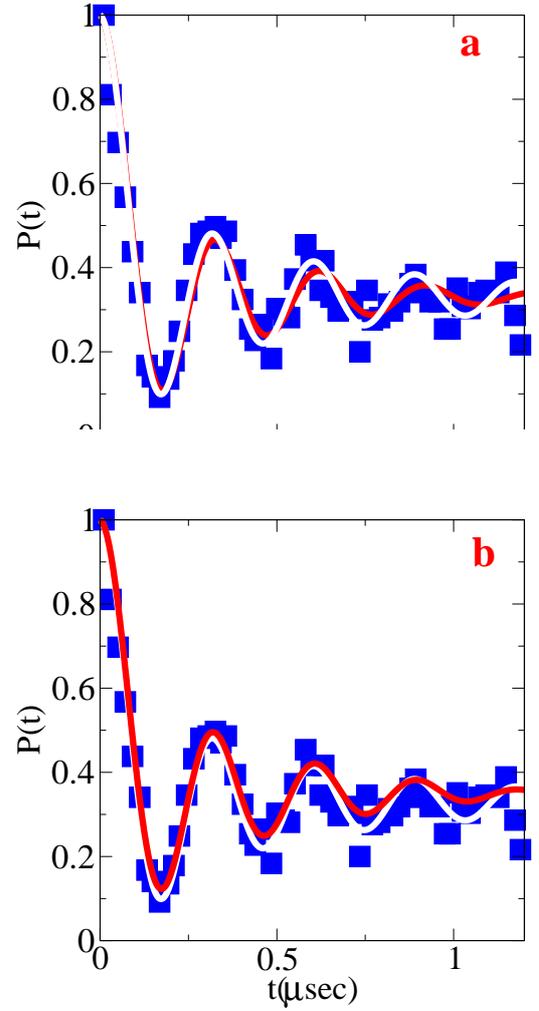

	\begin{center}
		\includegraphics[angle = 0, width=0.4\textwidth]{plot1m.eps}
		\includegraphics[angle = 0, width=0.4\textwidth]{plot2m.eps}
		\caption{Muon polarization vs time.
                  Blue squares  show T=5K data form   Ref.\cite{Keller2014}
      Panel a: The red curve shows the best fit of data with ``modified
      `spin stripe B'' model at $p=0.9$.
Panel b: The red curve shows the best fit of data with ``modified
      spin stripe C'' model at $p=0.9$.
The white line in both panels represents the fit at $p=1$ that is identical
to the spin spiral fit shown by the red line in Fig.\ref{fit}a.
      }
		\label{fit1}
	\end{center}
\end{figure}

{\bf Modified spin stripe B model.}
Let us consider a modification of the spin stripe B model. The shape is like
that in Fig.\ref{FSS}b but the ratio of the small spin over the large spin,
$p=|S_1|/|S_0|$, is arbitrary instead of $p=1/\sqrt{2}$. What range of the
ratio is consistent with data? In panel a of Fig.\ref{fit1}
the red line  presents the best for $p=0.9$ (the fitting parameters are
$\omega=23\times 10^6sec^{-1}$, $\lambda_T=1.0\times 10^6sec^{-1}$,
${U_m}=0.67$). For comparison the white line represents the
fit at $p=1$ that is identical
to the spin spiral fit shown by the red line in Fig.\ref{fit}a.
The red fit is not  bad, but still it is
getting out of the phase compared to the data at large t.
At $p=0.95$  the ``modified spin-stripe B'' model becomes consistent with
$\mu$SR data. Hence, the range consistent with data is
$0.95 \le p \le 1$.  

{\bf Modified spin stripe C model.}
Finally let us consider a modification of the spin stripe C model: the
shape is like that in Fig.\ref{FSS}c but the ratio of the small spin over
the large spin, $p=|S_2|/|S_1|$, is arbitrary instead
of $p=\frac{\cos(3\pi/8)}{\cos(\pi/8)}=0.41$. In panel b of Fig.\ref{fit1}
The red line  presents the best for $p=0.9$ (the fitting parameters are
$\omega=23\times 10^6sec^{-1}$, $\lambda_T=0.7\times 10^6sec^{-1}$,
${U_m}=0.65$). Again,  for comparison the white line represents the
fit at $p=1$ that is identical
to the spin spiral fit shown by the red line in Fig.\ref{fit}a.
The red fit is not  bad, but still it is somewhat off the data at large t.
At $p=0.95$  the ``modified spin-stripe C'' model becomes consistent with
$\mu$SR data. Hence, again the range consistent with experiment is
$0.95 \le p \le 1$.

Thus, the spin stripe configurations become consistent with $\mu$SR data only if
$0.95 < p  < 1$, practically  an abrupt change of the staggered spin
direction. This implies that that holes are practically fully localized.
In the commonly considered model B, Fig.\ref{FSS}b, this is localization on
sites 2,6,etc. Such strong localization is inconsistent with direct
measurements~\cite{Abbamonte2005} that indicate a very small amplitude
of charge modulation,
$\delta n \approx 0.03\cos\left(\frac{2\pi n}{4}+\phi\right)$.\\

{\bf In conclusion.} I analyze the available $\mu$SR data for
La$_{2-x}$Ba$_x$CuO$_4$ ($x=\frac{1}{8}$) and show that the data  are  perfectly
consistent with coplanar spin spiral lying in the CuO$_2$-plane.
The value of the spin in the spiral is $S=0.37\times\frac{1}{2}$. 
I also show that the data are inconsistent with spin stripes.\\

{\bf Acknowledgments}\\
I thank  Hugo Keller for communicating the data and relevant references.
I also thank Guniyat Khaliullin for important comments.

\end{document}